\documentclass[preprint,3p, number,twocolumn,sort&compress]{elsarticle}
\usepackage{xcolor}
\usepackage{graphicx}
\usepackage{xurl}
\usepackage[
    colorlinks=true,
    linkcolor=blue,
    citecolor=blue,
    urlcolor=blue
]{hyperref}

\usepackage{comment}
\usepackage{amssymb} 
\usepackage{booktabs}
\usepackage{siunitx}
\DeclareSIUnit{\cps}{cps}
\DeclareSIUnit{\sample}{S}
\DeclareSIUnit{\Vov}{\ensuremath{V_{OV}}}

\begin{document}

\begin{frontmatter}
\title{Characterisation of Commercially Available NUV-MT Silicon Photomultipliers}

\author[a]{T.~Avgitas}
\author[a]{Z.~Balmforth\corref{cor1}}
\ead{zoe.balmforth@uni-hamburg.de}
\author[a]{I.~Manthos}
\author[a,b]{K.~Nikolopoulos}
\author[a]{C.~Toukmenidis}
\cortext[cor1]{Corresponding author}
\affiliation[a]{organization={Institute for Experimental Physics, University of Hamburg},
postcode={22761},
city={Hamburg},
country={Germany}}
\affiliation[b]{organization={School of Physics and Astronomy, University of Birmingham},
postcode={B15 2TT},
city={Birmingham},
country={United Kingdom}}

\begin{abstract}
Silicon Photomultipliers (SiPMs) based on NUV-MT technology offer improved photo-detection efficiency and reduced correlated noise compared to earlier designs, making them increasingly attractive for particle and astroparticle physics applications. We present a systematic characterisation of commercially available Broadcom AFBR-S4N series NUV-MT SiPMs over a wide range of temperatures and overvoltages, from \qty{-85}{\celsius} to \qty{23}{\celsius} and from \qtyrange{10}{16}{\volt} overvoltage ($\mathrm{V_{OV}}$). Key performance parameters are measured, including breakdown voltage, gain, signal-to-noise ratio, dark count rate, afterpulsing probability, and both internal and external optical crosstalk. At \qty{12}{\Vov} and \qty{23}{\celsius} a dark noise of \qty{107.2 +- 2.1}{\kilo\cps\per\square\milli\metre} is measured, consistent with manufacturer specifications. At \qty{12}{\Vov} and \qty{-30}{\celsius} the measurements yielded an average dark noise of \qty{1.69 +- 0.12}{\kilo\cps\per\square\milli\metre}, an average gain of \num{6.88 +- 0.15 e6}, an average signal to noise ratio of 13.6 $\pm$ 0.9, an average afterpulsing probability of \qty{0.59 +- 0.06}{\percent}, an average direct crosstalk probability of \qty{26.49 +- 0.28}{\percent}, an average delayed crosstalk probability of \qty{0.72 +- 0.10}{\percent}, and an average external crosstalk probability of \qty{0.10 +- 0.04}{\percent}.
\end{abstract}

\end{frontmatter}

\section{Introduction}
\label{sec:intro}

Silicon Photomultipliers (SiPMs) have been established as the photodetector 
technology of choice across a broad range of applications in particle and 
astroparticle physics, medical imaging, and photon-sensitive instrumentation~\cite{Piemonte:2019kll, Acerbi:2019qgp,Simon:2018xzl, Gundacker:2020cnv,BISOGNI2019118}. 
Compared to traditional photomultiplier tubes (PMTs), SiPMs offer increased single-photon 
sensitivity, high intrinsic gain, compact form factor, low operating voltage, 
insensitivity to magnetic fields, and superior mechanical 
robustness~\cite{Piemonte:2019kll,Acerbi:2019qgp,Gundacker:2020cnv,Klanner:2018ydn}.
However, the performance of SiPMs is strongly dependent on operating conditions, 
most notably temperature and bias voltage. 
Temperature fluctuations in particular can significantly affect 
detector performance: the gain, dark count rate, and correlated 
noise all vary strongly with temperature, degrading photon-counting 
accuracy and energy resolution if the operating conditions are not 
carefully controlled~\cite{Acerbi:2019qgp,Klanner:2018ydn}. This makes systematic performance characterisation over a wide range of operating 
conditions essential for detector design and operation.

The near-ultraviolet, metal-filled trench (NUV-MT) SiPM technology represents a significant advance over earlier SiPM designs, offering improved photon detection efficiency in the NUV range, reduced dark count rate (DCR), and lower correlated noise through the use of metal-filled optical isolation trenches between microcells~\cite{Merzi:2023}, while maintaining high gain and excellent single-photon resolution~\cite{Broadcom_AFBR}.
These properties make NUV-MT SiPMs particularly attractive for the detection of 
scintillation light in noble liquid detectors, time-of-flight measurements, 
and other high-precision photon-counting applications.

The growing adoption of SiPMs in experiments requiring cryogenic or 
thermally variable operating environments has driven extensive characterisation 
efforts across the community. Systematic studies of the temperature dependence 
of SiPM noise properties have been performed for devices from FBK, Hamamatsu, 
and SensL (now onsemi), spanning temperatures from room temperature down to liquid 
argon~\cite{Acerbi:2016ikf,Acerbi:2022xpt} and even millikelvin temperatures~\cite{QUEST-DMC:2025pqg}. 
These studies have established the qualitative picture of SiPM noise at low 
temperatures: dark count rate follows an Arrhenius-type suppression with 
decreasing temperature, after pulsing probability increases strongly at lower 
temperatures due to the Shockley-Read-Hall carrier release 
mechanism~\cite{Shockley1952}, and direct optical crosstalk probability 
remains largely independent of temperature. Low-temperature SiPM operation is of particular relevance for 
next-generation noble liquid detectors such as 
DarkSide-20k, currently under construction ~\cite{DarkSide-20k:2017zyg, Manthos:2023swh, DarkSide-20k:2025avf, DarkSide-20k:2024usz, DarkSide-20k:2026cdf},
and DUNE~\cite{DUNE:2020lwj}, where SiPMs are deployed as 
photon sensors in liquid argon at \qty{87}{\kelvin}.
Moderate cooling is also of direct relevance to high-energy physics 
tracking detectors: the LHCb SciFi tracker operates its SiPMs at 
\qty{-40}{\celsius} to suppress radiation-induced dark count rate, and 
further cooling to cryogenic temperatures is under investigation for 
the LHCb Upgrade~II~\cite{Curras-Rivera:2025clp}.

Despite the widespread characterisation of SiPM devices from other manufacturers~\cite{Acerbi:2016ikf, Gallina:2019fxt, Gola:2019idb, Acerbi:2022xpt}, the temperature-dependent characterisation of Broadcom NUV-MT devices remains limited, although these commercially available devices have been used in simulation studies~\cite{Pena-Rodriguez:2024vxh}. Recent studies of devices in the same NUV-MT series have begun to address low-temperature performance. 
For example, in Ref.~\cite{Niu:2025huc} the AFBR-S4N44P164M array has been characterised at temperatures between \qty{77}{\kelvin} and \qty{238}{\kelvin} and in  Ref.~\cite{Liu:2025csi} the AFBR-S4N66P024M dual-channel device, along with SiPMs from Hamamatsu and NDL, were characterised in the temperature range from \qty{30}{\kelvin} to \qty{293}{\kelvin}. In both cases the primary motivation was the potential application to cryogenic detectors, such as CsI-based CE$\nu$NS detectors. 

Furthermore, external optical crosstalk, wherein photons produced during an avalanche propagate to a neighbouring SiPM and trigger a secondary avalanche, remains sparsely studied in the literature~\cite{Gallacher:2025jyl, Li:2024jdq}. Existing measurements for NUV-MT devices have been performed in scintillator-coupled configurations~\cite{KRATOCHWIL2025116782}, where the measured quantity is strongly dependent on detector geometry. However, the intrinsic device-level external crosstalk for this device series has not been reported.

In this article, a systematic characterisation of commercially 
available Broadcom AFBR-S4N series NUV-MT 
SiPMs over a wide range of temperatures, \qty{-85}{\celsius} to \qty{23}{\celsius}, 
and operating voltages, \qtyrange{10}{16}{\volt} overvoltage ($V_{OV}$), is presented. Key performance parameters 
are measured, including breakdown voltage $V_{bd}$, gain, signal-to-noise ratio (SNR), 
correlated and uncorrelated internal noise components, such as dark noise, afterpulsing probability, and internal and external optical crosstalk. 

This article is organised as follows. Section~\ref{sec:sipms} 
describes the relevant SiPM physics and the devices under test. 
Section~\ref{sec:setup} presents the experimental setup and data acquisition 
procedure. Sections~\ref{sec:operational} and~\ref{sec:noise} present the 
results of the operational characteristics and noise characterisation 
measurements, respectively. Conclusions are drawn in 
Section~\ref{sec:conclusions}.

\section{Silicon Photomultipliers}
\label{sec:sipms}

SiPMs consist of arrays of Single Photon Avalanche Diodes (SPADs) operated in Geiger mode, allowing single-photon detection with excellent timing resolution. Each SPAD in the active area is biased above its breakdown voltage such that the absorption of a single photon, generating a primary electron-hole pair, is sufficient to initiate a self-sustaining avalanche through impact ionisation. The avalanche is subsequently quenched by a series resistor integrated into each SPAD, enabling stable avalanche operation and fast microcell recovery. The overall output signal from the SiPM is the sum of the responses from all fired microcells, enabling highly precise photon counting. 
A detailed overview of SiPM operating principles, performance parameters, and characterisation methods can be found in Refs.~\cite{Piemonte:2019kll, Klanner:2018ydn, Acerbi:2019qgp}

\subsection{SiPM noise}
\label{subsec:sipm_noise}

SiPM noise sources can be classified as either uncorrelated or correlated with respect to an earlier avalanche. Both categories exhibit strong dependences on temperature and operating voltage.

Dark count rate is the primary uncorrelated source of SiPM noise, arising from thermally-generated carriers within the silicon lattice which trigger avalanches in the absence of incident single photons. These signals are indistinguishable from single photon signals, thus contributing to the intrinsic noise of the SiPM. 

Afterpulsing, a source of correlated noise within SiPMs, occurs when charge carriers from initial avalanches become temporarily trapped in defects within the silicon. These carriers are subsequently released, triggering secondary avalanches within the same microcell on a timescale governed by the Shockley--Read--Hall emission mechanism~\cite{Shockley1952,Acerbi:2019qgp,Klanner:2018ydn}. Since the microcell has not yet fully recharged, the resulting afterpulse amplitude falls below an integer multiple of the single photoelectron (PE) amplitude, making afterpulses easily identifiable by their characteristic sub-1PE amplitude. 

Optical crosstalk, an additional source of correlated SiPM noise, arises when photons produced during an avalanche in one microcell propagate to neighbouring microcells, initiating additional avalanches~\cite{Acerbi:2019qgp,Klanner:2018ydn}. 
Crosstalk can be further classified as internal or external. Internal crosstalk occurs between microcells within the same SiPM, whereas external crosstalk occurs between microcells of neighbouring SiPMs~\cite{Boulay:2022rgb,Gallacher:2025jyl,McLaughlin:2021xat}. The NUV-MT technology suppresses internal crosstalk through the use of metal-filled optical isolation trenches between microcells, which absorb secondary photons before they can propagate to neighbouring microcells~\cite{Merzi:2023}. 
External crosstalk, dependent on the geometry and distance between neighbouring SiPM devices, has been identified as a potentially significant noise source in densely packed detector configurations~\cite{Boulay:2022rgb, Gibbons:2023iux, Gallacher:2025jyl, Li:2024jdq}.

Together, these sources of noise degrade the SiPM performance, making precise characterisation of their temperature and voltage dependencies essential for accurate photon detection in high-precision applications.

\subsection{The AFBR-S4N SiPM series}
\label{subsec:broadcom_sipms}

The SiPMs characterised in this paper are part of Broadcom's NUV-MT SiPM series, designed for high photon detection efficiency in the near-ultraviolet (NUV) range and optimised for low correlated noise while maintaining high gain and excellent single-photon resolution~\cite{Broadcom_AFBR}. 

Two models from this series are used in this work: the AFBR-S4N66P014M, a single-channel SiPM with a \qtyproduct{6 x 6}{\milli\metre} active area comprising \num{22428} microcells at a \qty{40}{\micro\metre} SPAD pitch, and the AFBR-S4N66P024M, a dual-channel SiPM comprising two such areas arranged side by side at a fixed SiPM pitch of \qty{0.86}{\milli\metre}. The single-channel model is used for the full operational and noise characterisation campaign, while the dual-channel model is employed specifically to study external optical crosstalk between neighbouring SiPMs. Both share the same underlying microcell design and key performance parameters, as specified by the manufacturer and summarised in Table~\ref{tab:sipm_specs}.
The SiPMs are mounted on a custom PCB to provide power and signal readout.

\begin{table}[h!]
    \centering
    \resizebox{\columnwidth}{!}{%
    \begin{tabular}{lc}
        \toprule
        \textbf{Parameter} & \textbf{Value} \\
        \midrule
        Breakdown voltage, $V_{bd}$ & \qty{32.5}{\volt} \\
        Temperature coefficient of $V_{bd}$ & \qty{30}{\milli\volt\per\celsius} \\
        Dark noise ($R_{DN}$) & \qty{125}{\kilo\cps\per\square\milli\metre} \\
        Gain & \num{7.3e6} \\
        Optical crosstalk probability, $p_{CT}$ & \num{0.23} \\
        Afterpulse probability, $p_{AP}$ & \num{<0.01} \\
        Recharge time coefficient & \qty{55}{\nano\second} \\
        \bottomrule
    \end{tabular}%
    }
    \caption{Key performance parameters of the AFBR-S4N series SiPMs, 
    as specified by Broadcom 
    at \qty{12}{\Vov} and \qty{25}{\celsius}~\cite{Broadcom_AFBR}.}
    \label{tab:sipm_specs}
\end{table}

\section{Experimental setup}
\label{sec:setup}

\subsection{SiPM Characterisation Setup}
\label{subsec:setup_sipm_char}

\begin{figure}[h!]
    \centering
    \includegraphics[width=0.95\linewidth]{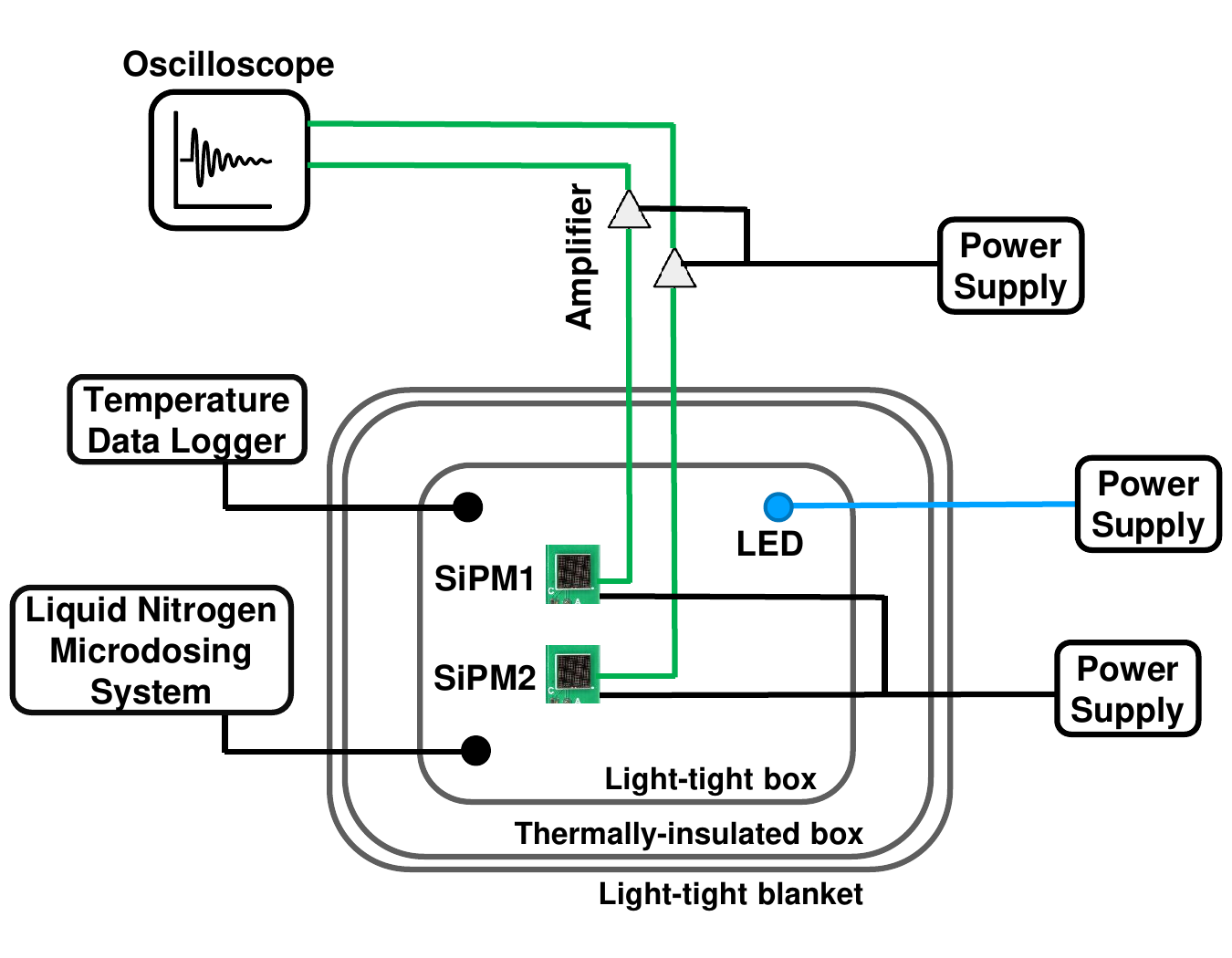}
    \caption{Schematic of the experimental setup used for the characterisation measurements.
    \label{fig:setup}}
\end{figure}

The characterisation measurements are performed in an ISO-7 certified 
clean room at the University of Hamburg. Two AFBR-S4N66P014M SiPMs are simultaneously characterised, 
mounted inside a light-tight box enclosed within a thermally insulated 
box and wrapped with a light-tight blanket to achieve the dark conditions 
necessary for accurate noise rate measurements. A schematic of the 
experimental setup is shown in Fig.~\ref{fig:setup}.

SiPM bias is provided by a Keithley 2231A-30-3 power supply, with a  Keithley 2450 Source Meter Unit  
used for IV curve acquisitions, where greater precision in the measured current draw is required. The SiPM output signal is amplified using a Mini-Circuits ZFL-1000LN+ low-noise amplifier before being digitised by a TektronixMSO44B  oscilloscope operating at \qty{1}{\giga\hertz} bandwidth and \qty{3.125}{\giga\sample\per\second} sampling rate. An LED with 
a peak wavelength of \qty{470}{\nano\metre}, 
is mounted inside the light-tight box and used to illuminate the SiPMs for IV curve  acquisitions.

A stable, temperature-controlled environment is achieved using a 
Norhof LN$_2$ Microdosing System, which supplies liquid nitrogen to the 
internal light-tight box and maintains the desired temperature within \qty{\pm1}{\celsius}. Temperature is 
continuously monitored using a Type~T Class~1 thermocouple and a data logger. 

\subsection{Data Acquisition}
\label{subsec:daq}

Data were acquired for SiPM overvoltages of \qtylist{10;12;14;16}{\Vov} 
at temperatures ranging from \qty{-85}{\celsius} to \qty{23}{\celsius}. 
At each temperature, an IV curve is acquired to determine the 
breakdown voltage $V_{bd}$, followed by noise data acquisition for each overvoltage.

For the IV curve measurement, the SiPM bias voltage is incremented and the current drawn recorded. The breakdown voltage $V_{bd}$ 
is estimated as the voltage at which the derivative of the logarithmic 
current with respect to voltage, $\mathrm{d\ln I/dV}$, is maximised.

During the noise data acquisition, the SiPM bias voltage $V_{bias}$ is set 
according to

\begin{equation}
    V_{bias} = V_{bd} + V_{OV}. 
    \label{eq:bias}
\end{equation}

\noindent Noise data are configured with a trigger threshold set below the single PE amplitude, and individual waveforms are recorded. For each temperature and overvoltage, a total acquisition window of \qty{200}{\milli\second} is used. 
The data are processed offline to extract the SiPM characterisation parameters.

\subsection{External crosstalk configuration}
\label{subsec:setup_exCT}

For the external crosstalk measurements, two AFBR-S4N66P024M 
dual-channel SiPMs are installed in the same experimental setup as described in Section~\ref{subsec:setup_sipm_char}, but separated by an opaque barrier to ensure optical isolation between the two devices. The LED is mounted within the barrier, with apertures on each side allowing simultaneous illumination of both SiPMs. 
Noise data are acquired using \qty{100}{\micro\second} frames at \qty{-30}{\celsius} for overvoltages of \qtylist{12;14;16}{\Vov}.
 
\section{Operational Characteristics}
\label{sec:operational}

The SiPM characterisation results are presented over a range of operating 
temperatures and overvoltages. Measurements of breakdown voltage, pulse shape, 
gain, and signal-to-noise ratio are discussed in the following subsections.

\subsection{Breakdown Voltage}
\label{subsec:vbd}

\begin{figure}[b!]
    \centering
    \includegraphics[width=0.95\linewidth]{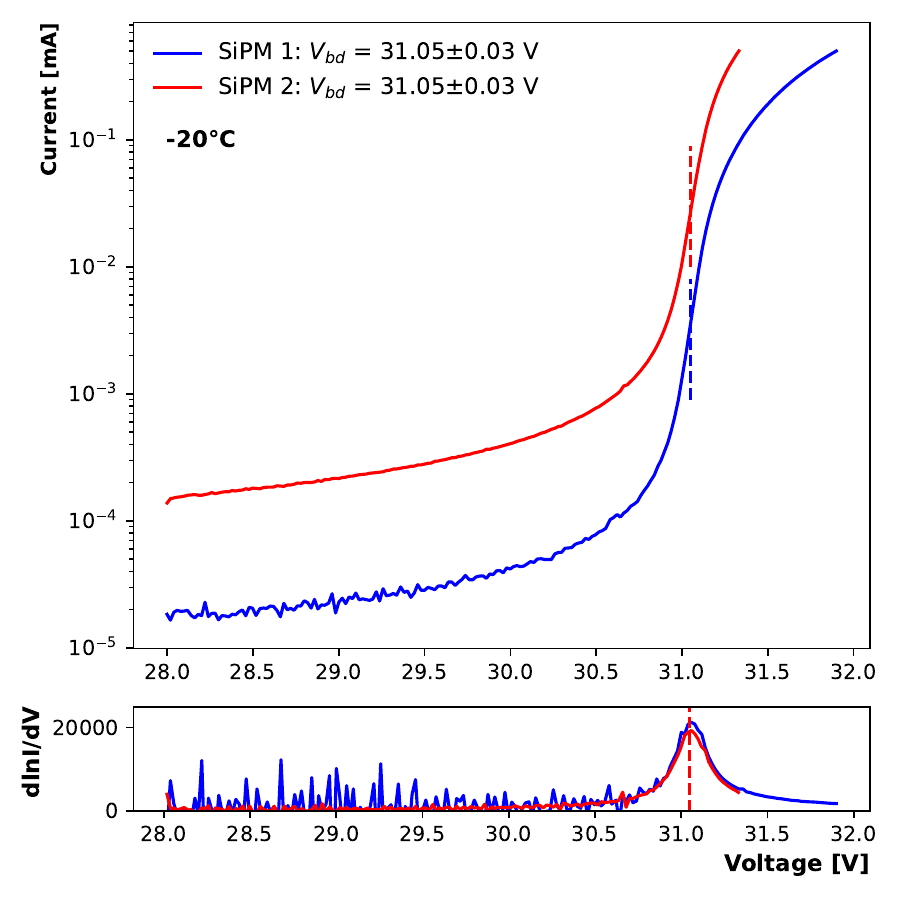}
    \caption{IV curve obtained at \qty{-20}{\celsius} for both SiPMs. The $V_{bd}$, estimated from the maximum of $\mathrm{d\ln I/dV}$, is indicated by the dashed lines.
    \label{fig:vbd_calculation}}
\end{figure}

The breakdown voltage of a SiPM depends on temperature through its effect on the carrier multiplication probability. Higher temperatures increase phonon scattering and reduce carrier mobility, effectively requiring a larger bias voltage to trigger an avalanche \cite{Klanner:2018ydn,Acerbi:2019qgp}. The IV curves of two SiPMs at \qty{-20}{\celsius} are shown in the top panel of Fig.~\ref{fig:vbd_calculation}, together with the derivative $\mathrm{d\ln I/dV}$ shown in the bottom panel. The breakdown voltage is indicated by the dashed lines. 

\begin{figure}[h!]
    \centering
        \includegraphics[width=0.95\linewidth]{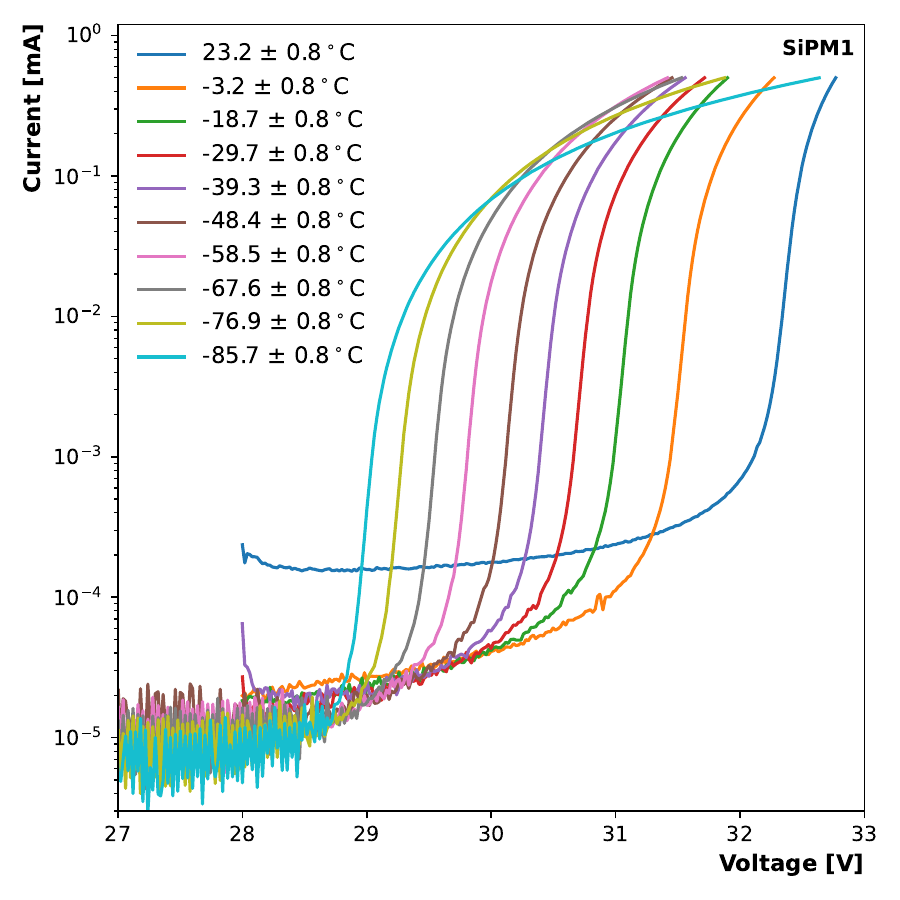}
    \caption{IV curves at temperatures from \qty{-85}{\celsius} to \qty{23}{\celsius} for SiPM 1. SiPM 2 yielded similar results.
    \label{fig:iv-curves}}
\end{figure}

\begin{figure}[h!]
    \centering
    \includegraphics[width=0.95\linewidth]{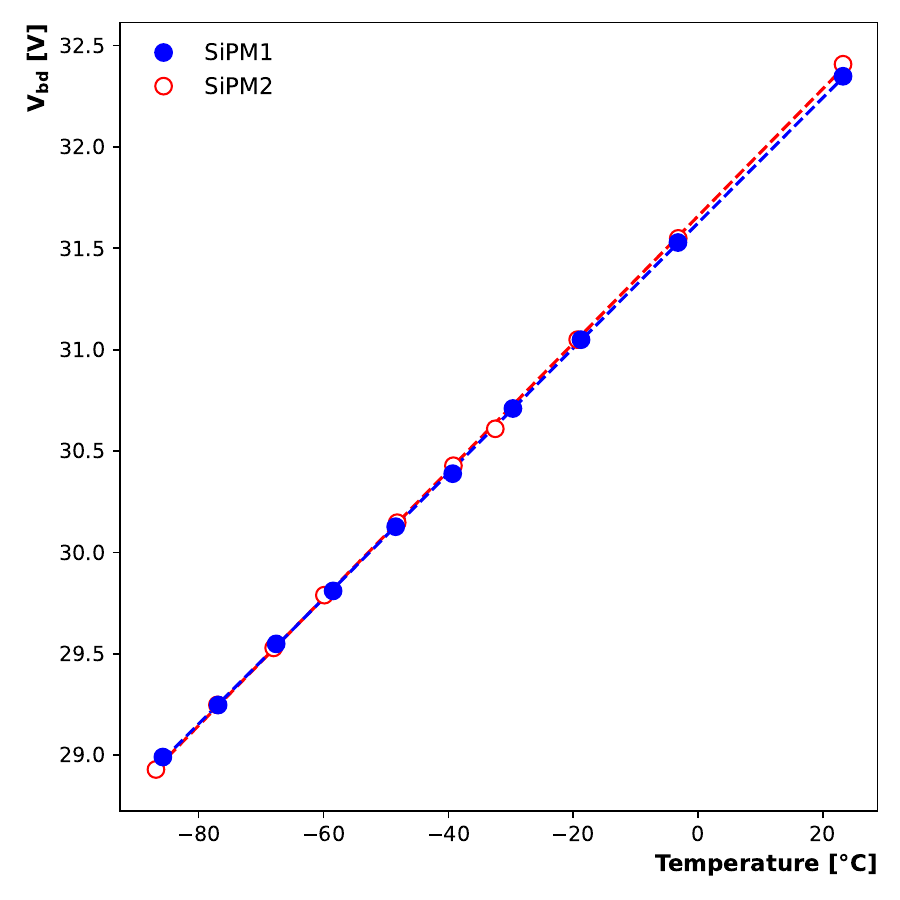}
    \caption{Breakdown voltage as a function of temperature for both SiPMs. Extracting the gradient from a combined linear fit of the two SiPMs results in a $V_{bd}$ dependence on temperature of \qty{31.2 +- 0.3}{\milli\volt\per\celsius}. Uncertainties are included in both axes but are not visible.
    \label{fig:Vbd_vs_temp}}
\end{figure}

IV curves acquired across the full temperature range are shown for one SiPM in Fig.~\ref{fig:iv-curves}. Figure~\ref{fig:Vbd_vs_temp} shows that the breakdown voltage exhibits a linear dependence on temperature. 
The temperature dependence of $V_{bd}$ extracted through a combined linear fit of both SiPMs is \qty{31.2 +- 0.3}{\milli\volt\per\celsius}. 
This is to be compared with the manufacturer specification of \qty{30}{\milli\volt\per\celsius}~\cite{Broadcom_AFBR}. The bias voltage $V_{bias}$ must therefore be adjusted at each temperature to maintain a consistent overvoltage, as described in Eq.~\ref{eq:bias}.

A recent characterisation of the AFBR-S4N66P024M at temperatures down to \qty{30}{\kelvin} reports a temperature coefficient of \qty{34.3}{\milli\volt\per\kelvin} above \qty{130}{\kelvin}~\cite{Liu:2025csi}, somewhat higher than the values measured here; the difference is likely attributable to the different $V_{bd}$ extraction method employed therein, namely $\sqrt{I}$--$V$ fitting rather than $\mathrm{d \ln I/dV}$ maximisation.

\subsection{Gain}
\label{subsec:gain}

Noise data acquired with the oscilloscope are processed offline to identify SiPM pulses within waveforms. The pulse amplitude, defined as the maximum height of the waveform above the local baseline within a sliding window of \qty{4.8}{\nano\second}, is used as the main pulse identification metric. This window may introduce dead time, during which a genuine dark noise pulse may go undetected. Assuming Poisson statistics with a mean equal to the dark noise rate at a given temperature, the probability that at least one pulse falls within this dead time window is calculated to be negligible for all temperatures considered in this characterisation campaign. 

\begin{figure}[h!]
    \centering
    \includegraphics[width=0.95\linewidth]{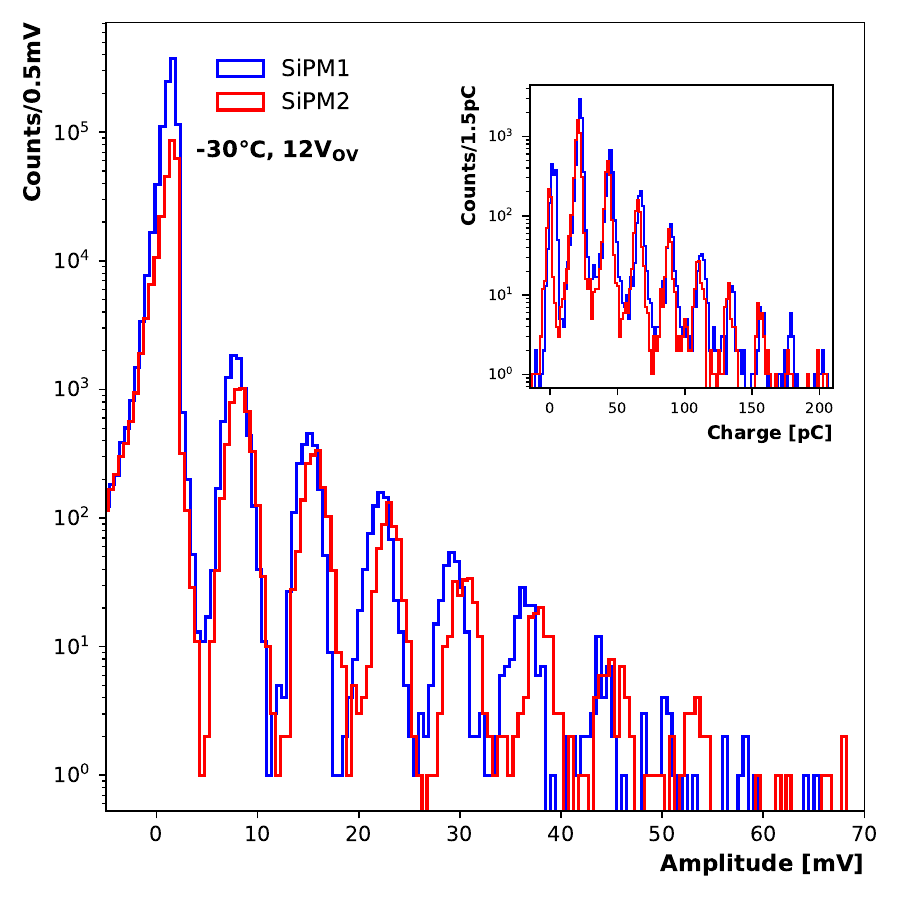}
    \caption{Single photoelectron (SPE) amplitude and charge spectra extracted from noise 
    data taken at \qty{-30}{\celsius} and \qty{12}{\Vov}. 
        \label{fig:fingerplot}}
\end{figure}

\begin{figure}[h!]
    \centering
    \includegraphics[width=0.95\linewidth]{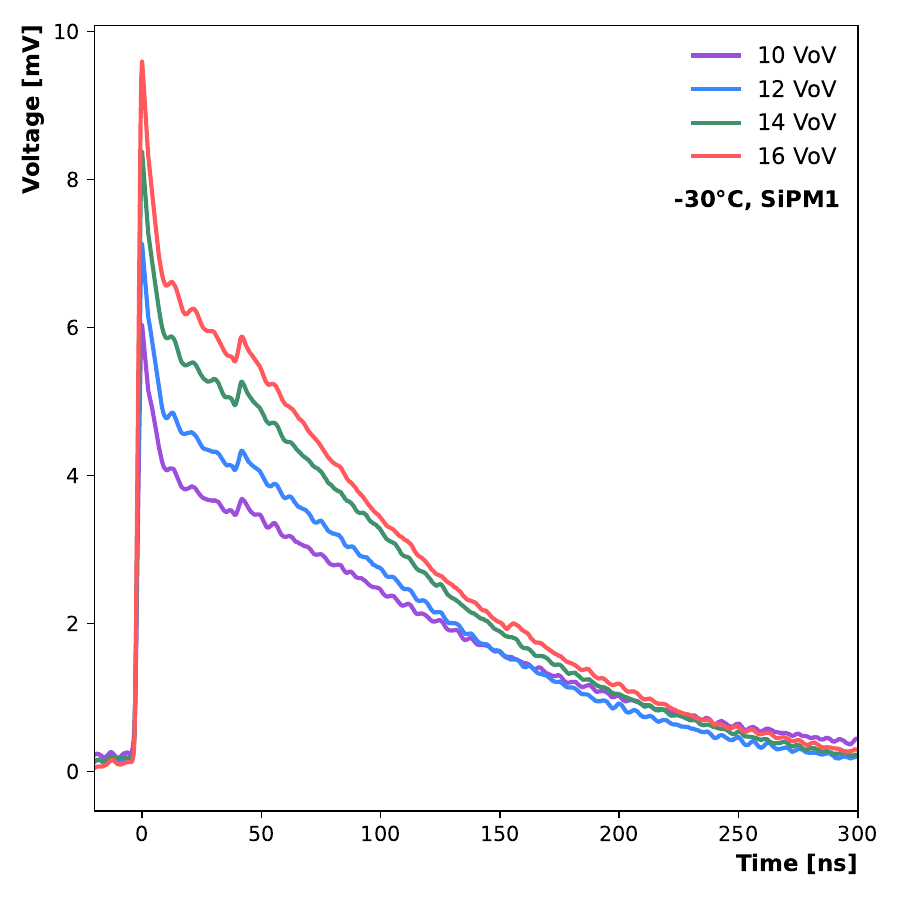}
    \caption{The average of \num{1000} single PE event waveforms for SiPM 1 at 
    \qty{-30}{\celsius} for overvoltages of \qtyrange{10}{16}{\Vov}. Afterpulses can clearly be seen occurring at \qty{50}{\nano\second} across all four overvoltage waveforms, and also at \qty{150}{\nano\second} for \qty{16}{\Vov}. 
    \label{fig:waveforms_n30_1pe}} 
\end{figure}

The pulse amplitude spectrum, obtained by generating a histogram of amplitudes 
of all identified pulses in a given acquired noise data run, reveals discrete peaks 
corresponding to signals from increasing numbers of photoelectrons, demonstrating the single-photon resolution of the SiPMs. This is commonly referred to as the single photoelectron (SPE) spectrum,  and is shown for both SiPMs at \qty{-30}{\celsius} and \qty{12}{\Vov} in Fig.~\ref{fig:fingerplot}. The average single PE pulse shape can be extracted from noise data, and is shown in Fig.~\ref{fig:waveforms_n30_1pe} for the full \qtyrange{10}{16}{\Vov} range.

\begin{figure}[h!]
    \centering
    \includegraphics[width=0.95\linewidth]{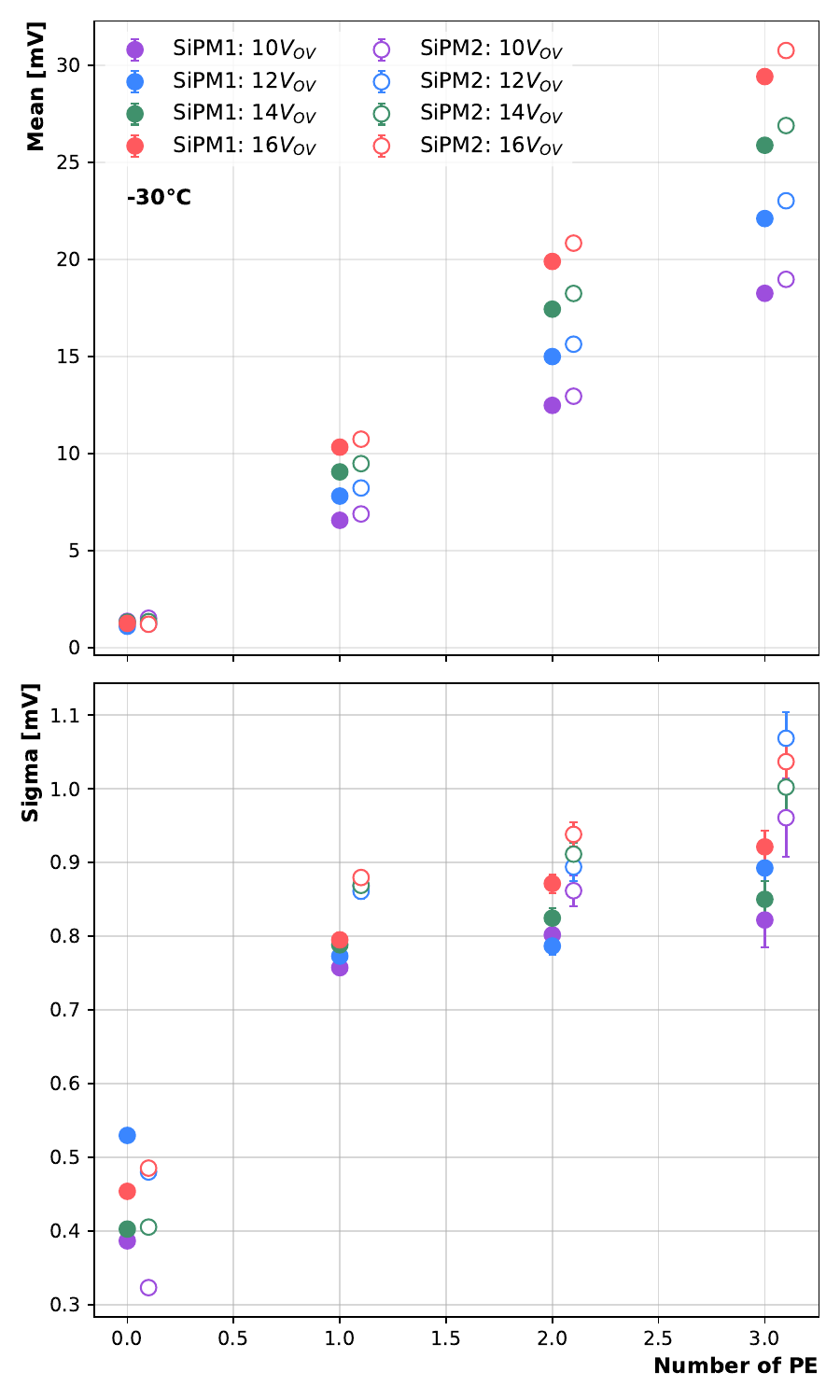}
    \caption{The mean and standard deviation extracted from Gaussian fits to the SiPM SPE amplitude spectra, shown as a function of the number of PEs at \qty{-30}{\celsius} for the full \qtyrange{10}{16}{\Vov} range. SiPM 2 data points have been shifted by \num{0.1} along the PE axis for visualisation purposes.
        \label{fig:mean_sigma_vs_pe} }
\end{figure}

Gaussian fits are performed to the first, second, and third peaks, as well as the pedestal centred around zero, to extract the mean amplitudes 
shown in Fig.~\ref{fig:mean_sigma_vs_pe}. The width of each peak reflects statistical fluctuations of the SPAD avalanche process, device-to-device variations, and electronic noise. 

The pulse charge spectrum, obtained by integrating all identified pulses over a window of \qty{300}{\nano\second}, is shown in the inset of~Fig.~\ref{fig:fingerplot}. The pulse charge can be used to calculate the SiPM gain, which is directly related to the separation between neighbouring peaks in the SPE spectrum. 
Since the SiPM output signal is amplified, 
the amplifier gain is accounted for when calculating the absolute SiPM gain. A dedicated calibration 
measurement yielded an amplifier gain of $A = \num{20.9 +- 0.6}$. 
The SiPM gain is then calculated as

\begin{equation}
    \label{eq:gain}
    G = \frac{Q_{{2}PE} - Q_{1PE}}{A\cdot R \cdot e},
\end{equation}

\noindent where $Q_{2\mathrm{PE}} - Q_{1\mathrm{PE}}$ denotes the charge difference between adjacent 2 and 1 photoelectron peaks, $A$ is the amplifier gain, $R$ is the impedance at the oscilloscope input, and $e$ is the elementary charge. 

\begin{figure}[h!]
    \centering
    \includegraphics[width=0.95\linewidth]{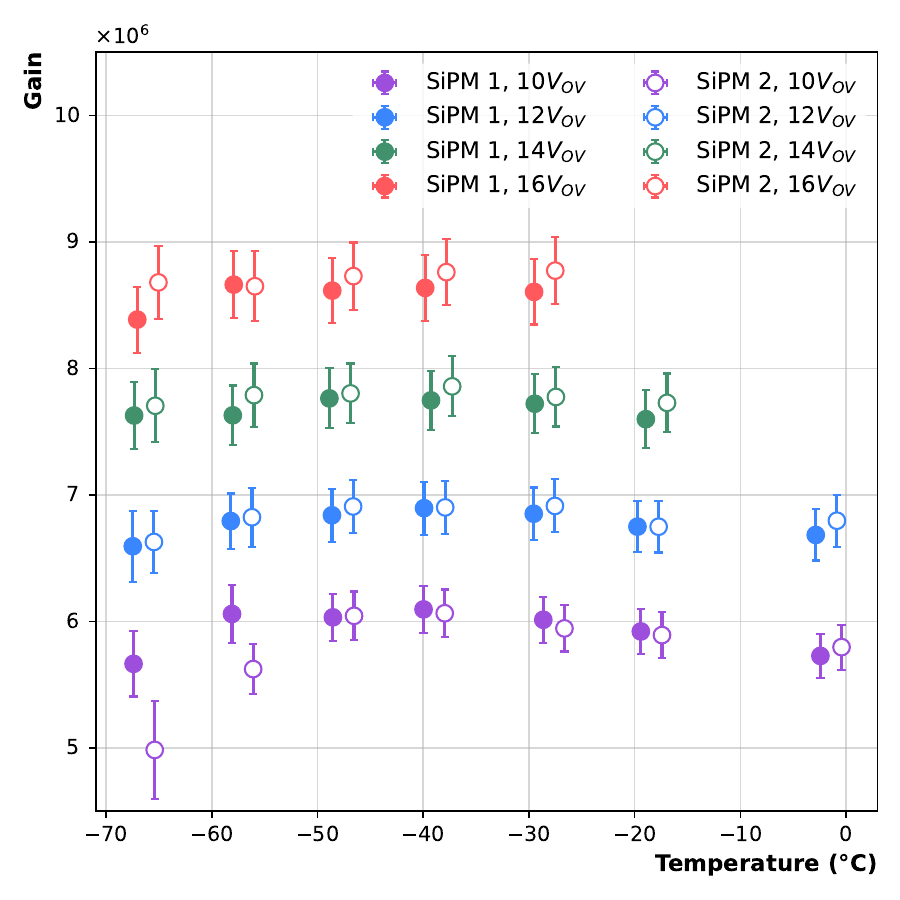}
    \caption{SiPM gain as a function of temperature for the full \qtyrange{10}{16}{\Vov} range, for both SiPMs. SiPM 2 data points have been shifted by \qty{2}{\celsius} along the temperature axis for visualisation purposes.
    \label{fig:gain_vs_temp}}
\end{figure}

The SiPM gain as a function of temperature for the full \qtyrange{10}{16}{\Vov} range is shown in Fig.~\ref{fig:gain_vs_temp}.
The gain increases with overvoltage, with the average gain across both SiPMs varying from \num{6.0 +- 0.1 e6} to \num{8.7 +- 0.2 e6} over the range \qtyrange{10}{16}{\Vov} at \qty{-30}{\celsius}, reflecting the larger avalanche charge at higher bias. Fluctuations in temperature correspond to variations in gain through the temperature dependence of $V_{bd}$, making stable operation dependent on either precise thermal control or active bias voltage compensation.

\subsection{Signal-to-Noise ratio}
\label{subsec:snr}

The SPE spectrum described in Section~\ref{subsec:gain} can be used to define a signal-to-noise ratio which characterises the ability to resolve SiPM signals above background noise. 
The SNR is defined as

\begin{equation}
    \label{eq:snr}
    SNR = \frac{\mu_{1PE} - \mu_{0PE}}{\sigma_{0PE}}.
\end{equation}

\noindent where $\mu_{1\mathrm{PE}}$ and $\mu_{0\mathrm{PE}}$ are the mean amplitudes of the 1~PE and pedestal peaks, respectively, and $\sigma_{0\mathrm{PE}}$ 
is the width of the Gaussian fit to the pedestal. A larger SNR indicates better separation power between photoelectron peaks and thus better single-photon resolution. 

\begin{figure}[h!]
    \centering
    \includegraphics[width=0.95\linewidth]{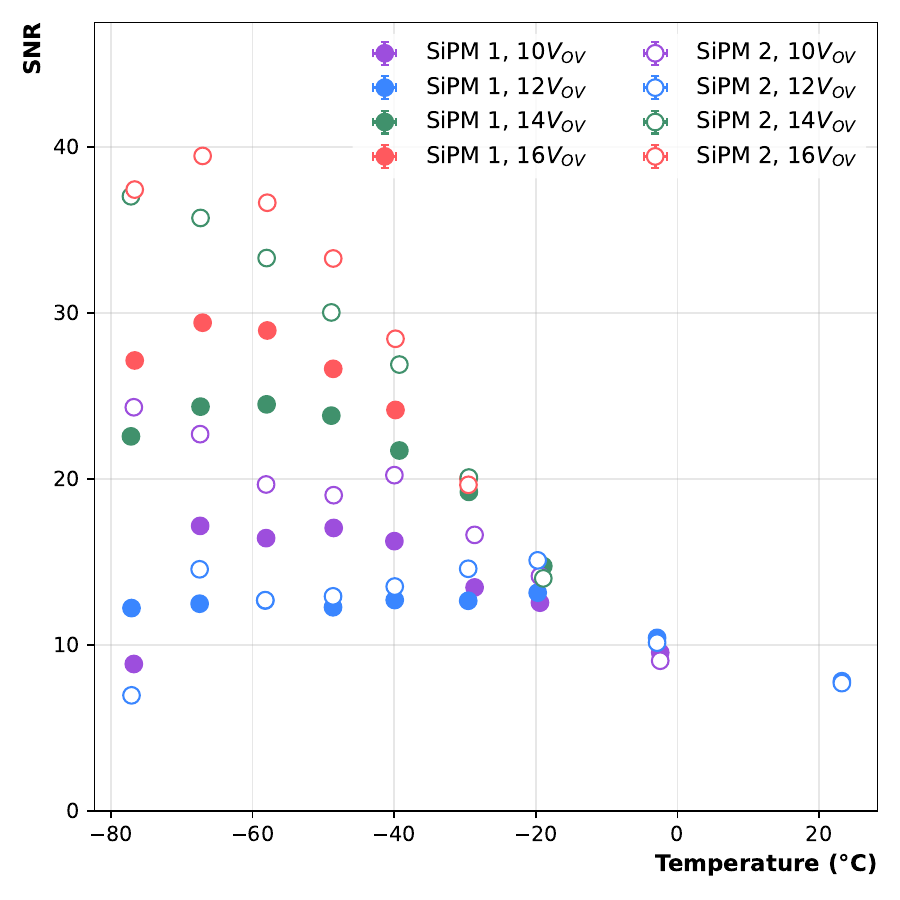}
    \caption{The SNR as a function of temperature for the full \qtyrange{10}{16}{\Vov} range for both SiPMs.}
    \label{fig:snr_v_temp}
\end{figure}

The SNR as a function of temperature, across the full \qtyrange{10}{16}{\Vov} range, is shown in Fig.~\ref{fig:snr_v_temp}.
SNR increases with decreasing temperature, varying from 7.8 $\pm$ 1.3 to 9.6 $\pm$ 2.6 over the range \qtyrange{+23}{-80}{\celsius} at \qty{12}{\Vov} due to the reduction in noise at lower temperatures. 
The SNR values also show the expected increase with overvoltage, varying from 15.1 $\pm$ 1.6 to 19.8 $\pm$ 0.2 over the range \qtyrange{10}{16}{\Vov} at \qty{-30}{\celsius}, reflecting the higher gain and increase in $\mu_{1PE}$ at increased bias voltage.

\section{Noise Characterisation}
\label{sec:noise}

The SiPM noise sources described in Section~\ref{subsec:sipm_noise} are characterised using an analysis method based on measuring the time differences between pulses in the acquired data, following the general characterisation framework discussed in Ref.~\cite{Klanner:2018ydn}. This method allows correlated and uncorrelated noise sources to be distinguished 
and their rates to be estimated independently. In the following, the waveform data are processed offline with an additional moving average filter to facilitate the analysis. 

\subsection{Time distribution method}
\label{subsec:deltat}
SiPM noise rates are estimated using the time distribution between selected primary pulses, and subsequent secondary pulses~\cite{Butcher:2017twm}. 

Primary pulses are required to have a 1~PE amplitude, with no pulse present 
in a preceding time window 
which is sufficiently large to minimise the 
probability of the primary pulse itself being a correlated noise 
event, whilst ensuring that any correlated pulses not associated 
with the primary pulse have a negligible probability of being 
selected as the secondary pulse. 

The selection of secondary pulses is less stringent: the first pulse above the baseline following a primary pulse is selected. 
Only the first subsequent pulse is considered, to avoid the statistical complications arising from 
correlated pulses of correlated pulses~\cite{Butcher:2017twm}.

\begin{figure}[h!]
    \centering
    \includegraphics[width=0.95\linewidth]{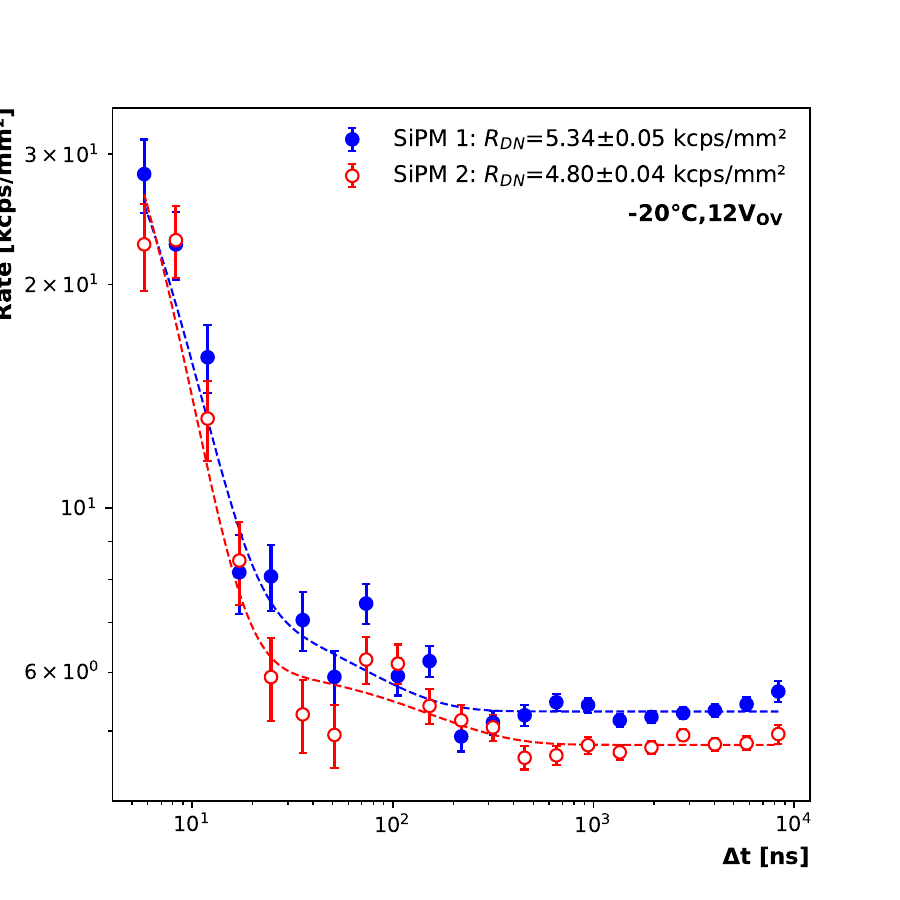}
    \caption{Corrected time distribution of secondary pulses relative 
    to primary pulses at \qty{-20}{\celsius} and 
    \qty{12}{\Vov}, following the unshadowing procedure of 
    Ref.~\cite{Butcher:2017twm}. The asymptotic behaviour at long 
    times gives the dark noise, while the excess at short times 
    reflects correlated noise contributions.
    \label{fig:dt_unshadowed}}
\end{figure}

Once a suitable population of pulses is identified, the measured distribution of time differences between secondary and primary pulses gives the probability, $p_i$, of a secondary pulse occurring at time $t_i$ after a primary pulse at time $t=0$:

\begin{equation}
\label{eq:prob_secondary_pulse}
    p_i = \frac{n_i}{N(t_{i+1} - t_i)}
\end{equation}

\noindent where $n_i$ is the number of secondary pulses observed in 
the time interval $[t_i, t_{i+1}]$ and $N$ is the total number of 
primary pulses selected. To account for shadowing of late processes in the time distribution measurement by early processes, 
the unshadowing method of Ref.~\cite{Butcher:2017twm} is applied, 
yielding the corrected time distribution shown in 
Fig.~\ref{fig:dt_unshadowed}. By modelling the resulting time distribution with a double exponential plus constant component fit, correlated and uncorrelated noise sources can be distinguished and estimated. At short times, this time distribution reflects the combined contribution of correlated and uncorrelated noise, while at long times it converges to the uncorrelated dark noise alone.

\begin{figure}[h!]
    \centering
    \includegraphics[width=0.95\linewidth]{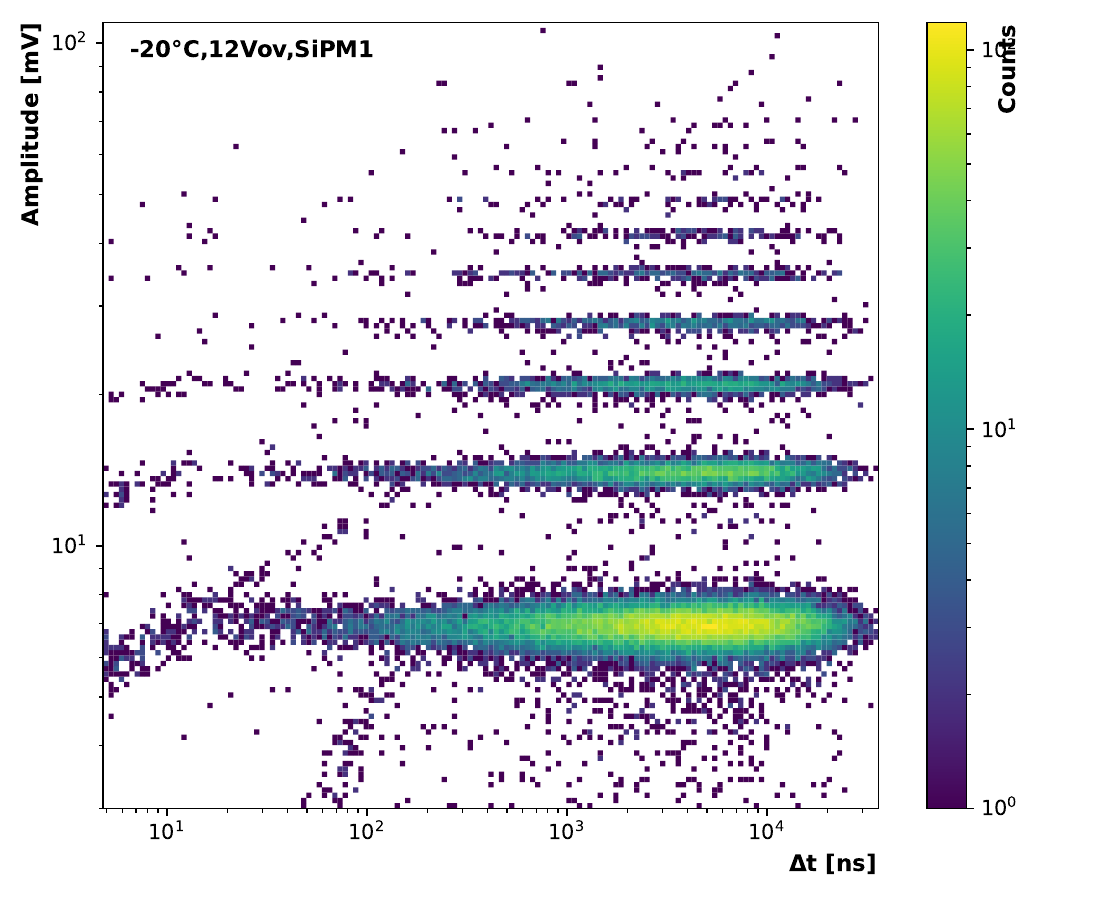}
    \caption{Amplitude as a function of $\Delta t$ between primary and secondary pulses at \qty{-20}{\celsius} and \qty{12}{\Vov}.\label{fig:amp_vs_dt}}
\end{figure}

The amplitude of identified secondary pulses as a function of their time difference $\Delta t$ from the primary pulse is shown in Fig.~\ref{fig:amp_vs_dt}, measured at \qty{-20}{\celsius} and \qty{12}{\Vov}. The discrete horizontal bands, which appear at integer multiples of the single PE amplitude, arise from signals corresponding to successive numbers of PE, also seen in the amplitude spectrum of Fig.~\ref{fig:fingerplot}. Several distinct correlated and uncorrelated noise populations can be distinguished in the pulse amplitude against time difference distribution.

At short time differences ($\Delta t \lesssim \qty{10}{\nano\second}$), secondary pulses with amplitudes consistent with integer multiples of the single PE amplitude can be attributed to delayed optical crosstalk~\cite{Gola:2019idb}, a source of correlated noise.

Afterpulses, another correlated noise source, can be identified at timescales ranging from $\Delta t \approx \qtyrange{10}{e3}{\nano\second}$ by their characteristic amplitude which falls below an integer multiple of the single PE amplitude~\cite{Butcher:2017twm,Garutti:2016hny}. 

The uncorrelated population with 1~PE amplitude at longer timescales ($\Delta t \gtrsim \qty{e3}{\nano\second}$) is attributed to primary dark counts, while the general population with amplitudes larger than 1~PE is consistent with noise correlated with these primary dark counts~\cite{McLaughlin:2021xat}: direct optical crosstalk. 

\subsection{Dark Noise}
\label{subsec:uncorrelated_noise}
	
Primary dark counts, arising from thermally-generated carriers triggering avalanches within individual SiPM microcells in the absence of incident photons, produce signals indistinguishable from single-photon events. The rate of these thermally induced avalanches is dependent on the silicon bandgap energy and the local electric field in each microcell, both of which are strongly temperature dependent. At higher temperatures, the phonon interaction probability increases, leading to a higher probability of charge carrier generation and, thus, an increase in primary dark count rate. 

Dark noise, $R_{DN}$, defined as the rate of primary dark counts and the associated direct crosstalk, is estimated using two independent methods: a counting method and the time distribution method described in Section~\ref{subsec:deltat}. In the counting method, the dark noise is estimated as

\begin{equation}
\label{eq:dn}
    R_{DN} = \frac{N_{0.5\,\mathrm{PE}}}{T} = R_{0.5\,\mathrm{PE}},
\end{equation}

\noindent where $N_{0.5\,\mathrm{PE}}$ is the total number of pulses with an amplitude exceeding 0.5~PE and $T$ is the total acquisition time. In the time distribution method, $R_{DN}$ is extracted from the asymptotic behaviour at long pulse time differences in Fig.~\ref{fig:dt_unshadowed}. Both methods yield consistent results across the full measurement range. 

\begin{figure}[h!]
    \centering
    \includegraphics[width=0.95\linewidth]{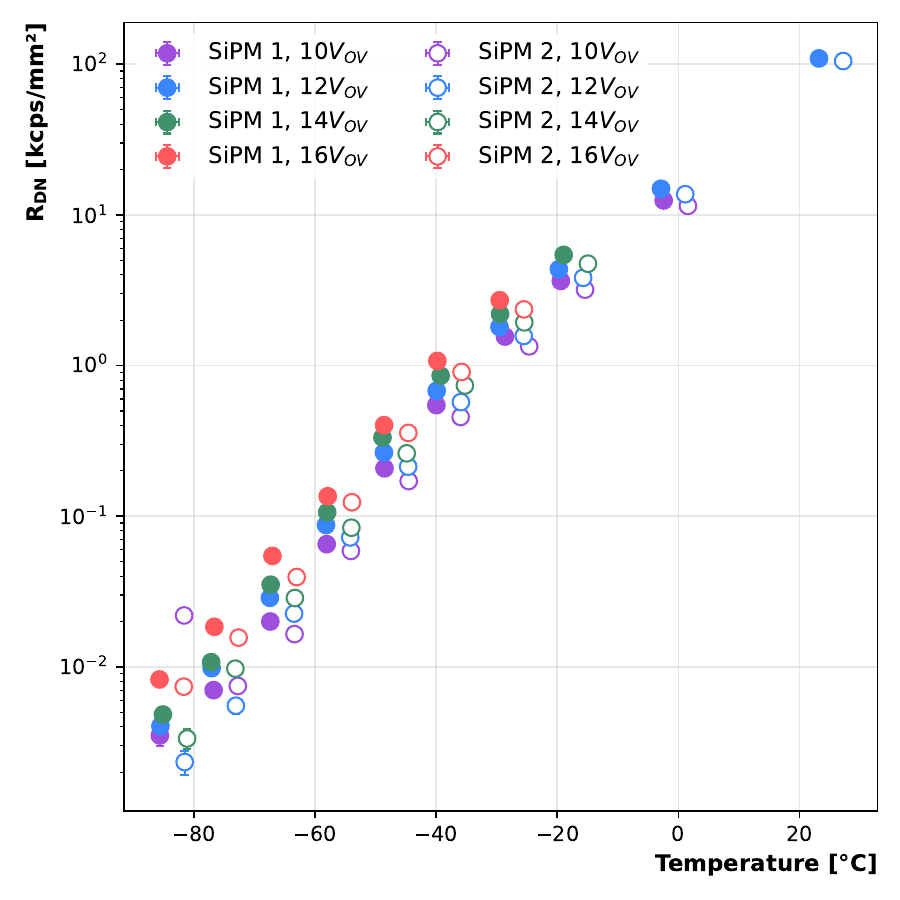}
    \caption{Dark noise as a function of temperature for the full \qtyrange{10}{16}{\Vov} range, for both SiPMs. The rate decreases by approximately four orders of magnitude between \qty{-85}{\celsius} and \qty{23}{\celsius}, consistent with thermally activated carrier generation. SiPM 2 data points have been shifted by \qty{4}{\celsius} along the temperature axis for visualisation purposes.}
    \label{fig:dcr_vs_temp}
\end{figure}

Dark noise follows an Arrhenius-type dependence on temperature, decreasing exponentially as the temperature is lowered~\cite{Acerbi:2019qgp,Klanner:2018ydn}; behaviour which is observed between \qty{-85}{\celsius} and \qty{23}{\celsius} in Fig.~\ref{fig:dcr_vs_temp} across the full \qtyrange{10}{16}{\Vov} range. For temperatures lower (higher) than \qty{-25}{\celsius}, dark noise is observed to change by a factor of 2 approximately every \qty{8}{\celsius} (\qty{6}{\celsius}). Both SiPMs exhibit good mutual agreement across the full temperature range.

The average dark noise across the two SiPMs is \qty{107.2 +- 2.1}{\kilo\cps\per\square\milli\metre} at \qty{12}{\Vov} and \qty{23}{\celsius}, consistent with the manufacturer specification of \qty{125}{\kilo\cps\per\square\milli\metre} at \qty{12}{\Vov} and \qty{25}{\celsius}~\cite{Broadcom_AFBR}, when the temperature discrepancy is accounted for. 

In Fig.~\ref{fig:dcr_vs_temp}, the dark count rate can also be seen to increase monotonically with overvoltage, rising from 
\qty{1.44 +- 0.11}{\kilo\cps\per\square\milli\metre} to 
\qty{2.54 +- 0.18}{\kilo\cps\per\square\milli\metre} over the range \qtyrange{10}{16}{\Vov} at \qty{-30}{\celsius}, consistent with the increasing Geiger discharge probability at higher bias. Both SiPMs exhibit consistent behaviour, with SiPM 2 showing a marginally lower rate across the \qtyrange{10}{16}{\Vov} range.

\subsection{Afterpulsing}
\label{subsec:correlated_noise}

Afterpulses, generated by charge carriers trapped in lattice defects within the same SPAD as the primary avalanche which are released and subsequently generate a secondary avalanche, typically occur on timescales ranging from tens of nanoseconds to microseconds 
after a primary pulse. As temperature decreases, the trap release time constant, governed by the Shockley--Read--Hall emission rate~\cite{Shockley1952}, increases, shifting the charge carrier release time distribution into the microcell recharge window, and thus increasing the probability of a secondary avalanche. 

\begin{figure}[h!]
    \centering
    \includegraphics[width=0.95\linewidth]{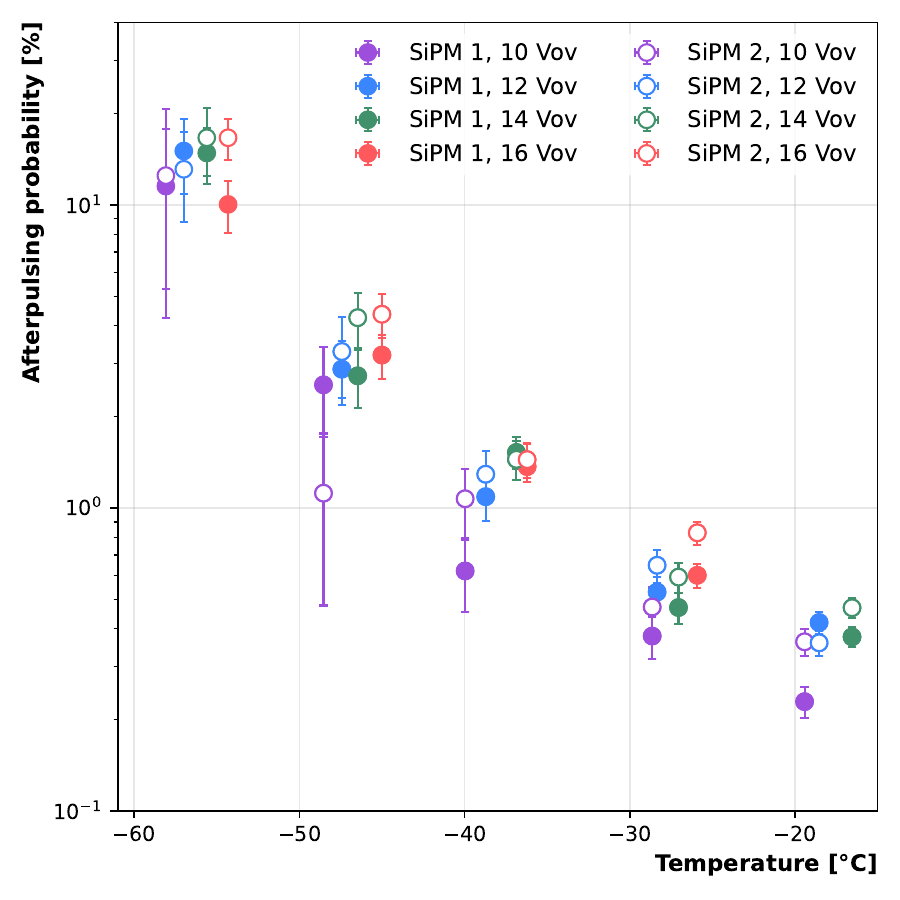}
    \caption{Afterpulsing probability as a function of temperature for the full \qtyrange{10}{16}{\Vov} overvoltage range. For visualisation purposes, data points for \qtyrange{12}{16}{\Vov} are incrementally shifted along the temperature axis by \qty{1.2}{\celsius}.}
    \label{fig:ap_vs_temp}
\end{figure}

The afterpulsing probability, $p_{AP}$, is estimated by integrating the secondary pulse rate over the relevant $\Delta t$ range in Fig.~\ref{fig:amp_vs_dt}, excluding contributions from dark noise and crosstalk. In Fig.~\ref{fig:ap_vs_temp}, $p_{AP}$ is presented across the full \qtyrange{10}{16}{\Vov} range, and a subset of temperatures where the probability of dark noise events occurring within the afterpulse $\Delta t$ selection window is negligible, to avoid shadowing the afterpulsing process in the selection of secondary pulses. The afterpulsing probability increases strongly with decreasing temperature, characteristic of thermally activated carrier release from trapping centres in the silicon bulk and at interfaces. The average $p_{AP}$ across the two SiPMs at \qty{12}{\Vov} increases from 
\qty{0.39 +- 0.03}{\percent} at \qty{-20}{\celsius} to \qty{14.1 +- 3.0}{\percent} 
at \qty{-58}{\celsius}.

The overvoltage dependence of $p_{AP}$ is also shown 
in Fig.~\ref{fig:ap_vs_temp}. The afterpulsing probability 
increases with overvoltage, consistent with the larger avalanche 
charge at higher bias increasing the number of carriers available 
for trapping at defect sites. The average probability for the two SiPMs increases from \qty{0.41 +- 0.05}{\percent} to \qty{0.71 +- 0.11}{\percent} 
over the range \qtyrange{10}{16}{\Vov} at \qty{-30}{\celsius}, in good agreement with the \qty{<1}{\percent} value reported by the manufacturer~\cite{Broadcom_AFBR} across the entire investigated overvoltage range. 

\subsection{Direct optical crosstalk}
\label{subsec:dct}

Direct optical crosstalk arises when photons emitted during an avalanche propagate to the active region of a neighbouring microcell, triggering a secondary Geiger discharge, as described in Section~\ref{subsec:sipm_noise}. The direct crosstalk probability, $p_{CT}$, is estimated from the ratio of dark noise hits above and below the 1.5~PE threshold:

\begin{equation}
    p_{CT} = \frac{R_{1.5\,\mathrm{PE}}}{R_{0.5\,\mathrm{PE}}},
    \label{eq:pct}
\end{equation}
\noindent where $R_{1.5\,\mathrm{PE}}$ and $R_{0.5\,\mathrm{PE}}$ 
denote the rates of pulses with amplitudes exceeding 1.5~PE and 
0.5~PE, respectively~\cite{Klanner:2018ydn}.

\begin{figure}[h!]
    \centering
    \includegraphics[width=0.95\linewidth]{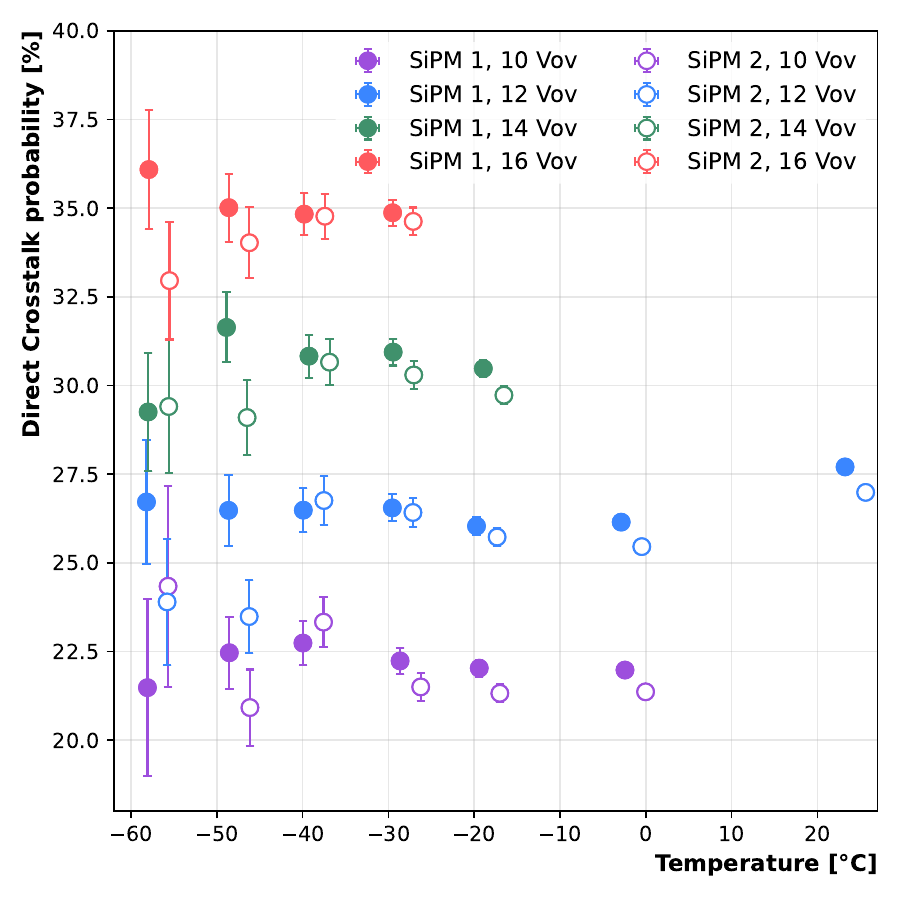}
    \caption{Direct crosstalk probability as a function of temperature for the full temperature range, for both SiPMs. The probability is approximately independent of temperature, consistent with the avalanche-driven crosstalk generation mechanism. SiPM 2 data points have been shifted by \qty{2.4}{\celsius} along the temperature axis for visualisation purposes.}
    \label{fig:DiCT_vs_temp}
\end{figure}

The temperature dependence of $p_{CT}$ is presented in Fig.~\ref{fig:DiCT_vs_temp} for the full \qtyrange{10}{16}{\Vov} range. For a given overvoltage, the crosstalk probability remains approximately constant over the measured temperature range, with the SiPM average varying between \qty{27.3 +- 0.4}{\percent} at \qty{23}{\celsius} and \qty{25.3 +- 1.4}{\percent} 
at \qty{-58}{\celsius} for \qty{12}{\Vov}. This approximate temperature independence is consistent with the expectation that the crosstalk generation mechanism is driven by the avalanche charge rather than by thermally generated carriers. Both SiPMs exhibit mutually consistent values within uncertainties across the full temperature range investigated.

The overvoltage dependence of $p_{CT}$ is also presented in Fig.~\ref{fig:DiCT_vs_temp} across the full temperature range. The crosstalk probability increases with overvoltage, varying from \qty{21.9 +- 0.4}{\percent} to \qty{34.8 +- 0.3}{\percent} 
over the range \qtyrange{10}{16}{\Vov} at \qty{-30}{\celsius}, consistent with the larger avalanche charge at higher bias enhancing the probability of secondary photon emission and subsequent crosstalk triggering in neighbouring microcells. While the values of $p_{CT}$ reported here exceed the manufacturer specification~\cite{Broadcom_AFBR} of \qty{23}{\percent} at \qty{12}{\Vov}, the observed overvoltage dependence and temperature independence of $p_{CT}$ is expected.

\subsection{Delayed optical crosstalk}
\label{subsec:deCT}

Delayed optical crosstalk arises when photons propagate to neighbouring microcells, generating carriers in the undepleted silicon bulk below the active region of the microcell. If these carriers diffuse into the microcell's active region before recombining in the silicon bulk, a delayed crosstalk event is generated~\cite{Gola:2019idb}. 

The delayed crosstalk probability, $p_{deCT}$, is governed by the diffusion length of the generated carriers. This diffusion length, which dictates the volume below the active region in which a photon induced carrier can be generated and successfully diffuse into the active region and trigger a delayed crosstalk event, increases with decreasing temperature as carrier lifetime is extended by reduced phonon scattering. 

\begin{figure}[h!]
    \centering
    \includegraphics[width=0.95\linewidth]{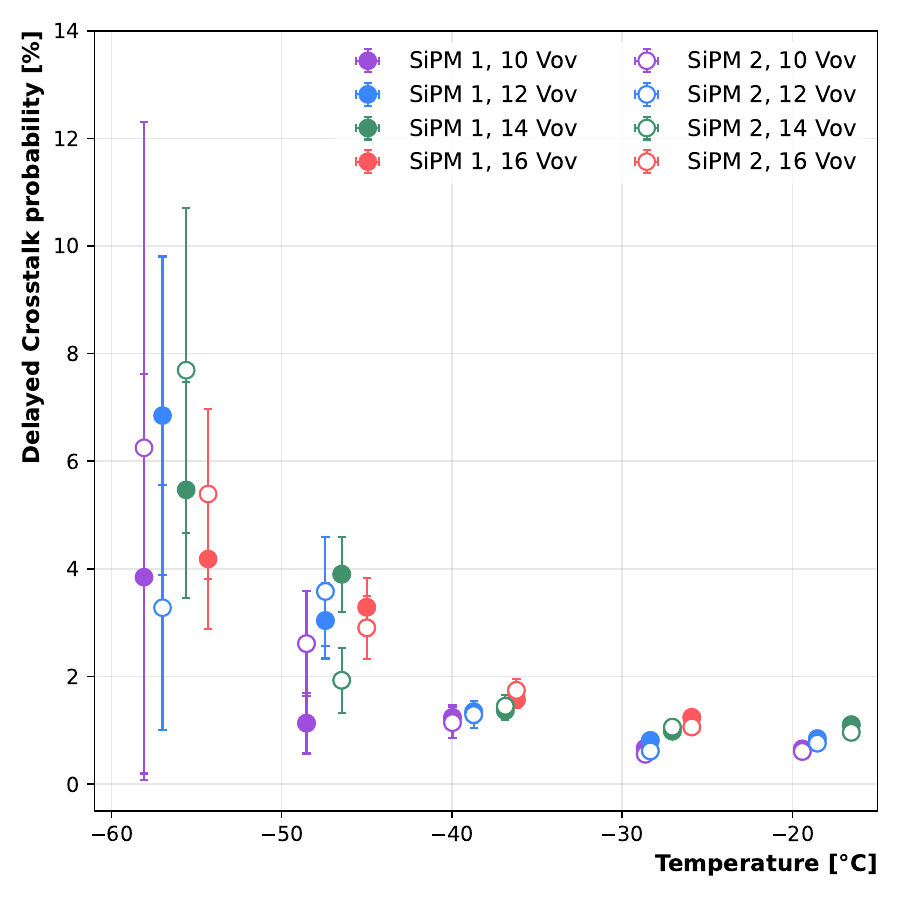}
    \caption{Delayed crosstalk probability as a function of overvoltage for a range of temperatures, for both SiPMs. For visualisation purposes, data points for \qtyrange{12}{16}{\Vov} are incrementally shifted along the temperature axis by \qty{1.2}{\celsius}.}
    \label{fig:DeCT_vs_ov}
\end{figure}

Figure~\ref{fig:DeCT_vs_ov} shows the delayed crosstalk probability across the full \qtyrange{10}{16}{\Vov} range for a subset of temperatures where the probability of dark noise events occurring within the delayed crosstalk $\Delta t$ selection window is negligible. The expected increase in probability at lower temperatures is seen across both SiPMs. The average $p_{deCT}$ across the two SiPMs at \qty{12}{\Vov} increases from \qty{0.80 +- 0.04}{\percent} at \qty{-20}{\celsius} to \qty{4.6 +- 1.8}{\percent} at \qty{-58}{\celsius}. 

Similarly to direct optical crosstalk, $p_{deCT}$ also typically increases with overvoltage since the larger avalanche charge at higher bias enhances the probability of secondary photon emission
and subsequent crosstalk triggering in neighbouring microcells. At \qty{-30}{\celsius}, the average $p_{deCT}$ across the two SiPMs increases from \qty{0.61 +- 0.06}{\percent} to \qty{1.15 +- 0.09}{\percent} over the range \qtyrange{10}{16}{\Vov}. 

\subsection{External crosstalk}
\label{subsec:exCT}

Unlike internal crosstalk, external crosstalk depends on the geometry and separation of neighbouring SiPMs~\cite{Boulay:2022rgb, Gibbons:2023iux, Gallacher:2025jyl, Li:2024jdq}. As described in Section~\ref{subsec:setup_exCT}, a dedicated measurement campaign was performed using two dual-channel AFBR-S4N66P024M devices, which each comprise of two \qtyproduct{6 x 6}{\milli\metre} SiPMs side by side at a fixed separation of \qty{0.86}{\milli\metre}.

\begin{figure}[h!]
    \centering
    \includegraphics[width=0.95\linewidth]{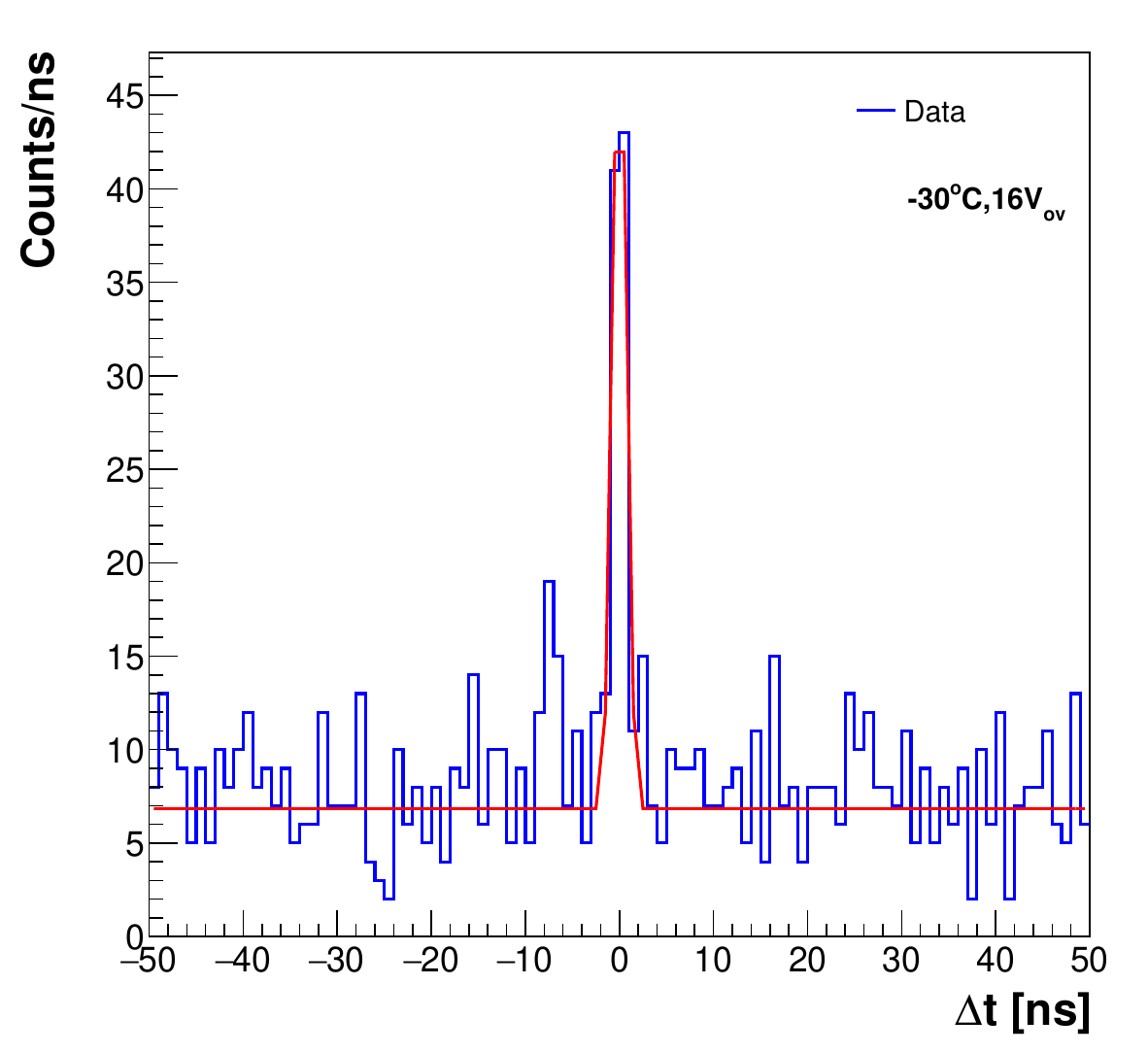}
    \caption{The $\Delta t$ distribution between the two channels of a double SiPM at \qty{-30}{\celsius} and \qty{16}{\Vov}. The excess of events at small time differences is consistent with the presence of external crosstalk.}
    \label{fig:exCT_deltaT}
\end{figure}

Noise data are acquired simultaneously from both channels of the double SiPM operated in dark conditions. 
For each pulse identified on the SiPM channel designated as the trigger, the closest in time pulse in the other SiPM channel is used to construct the inter-arrival time distribution, $\Delta t$, between the two SiPMs.
Figure~\ref{fig:exCT_deltaT} shows the resulting distribution for one of the double SiPMs at \qty{-30}{\celsius} and \qty{16}{\Vov}. The clear excess of events visible at small time differences is attributed to external crosstalk between SiPMs.

\begin{figure}[h!]
    \centering
    \includegraphics[width=0.95\linewidth]{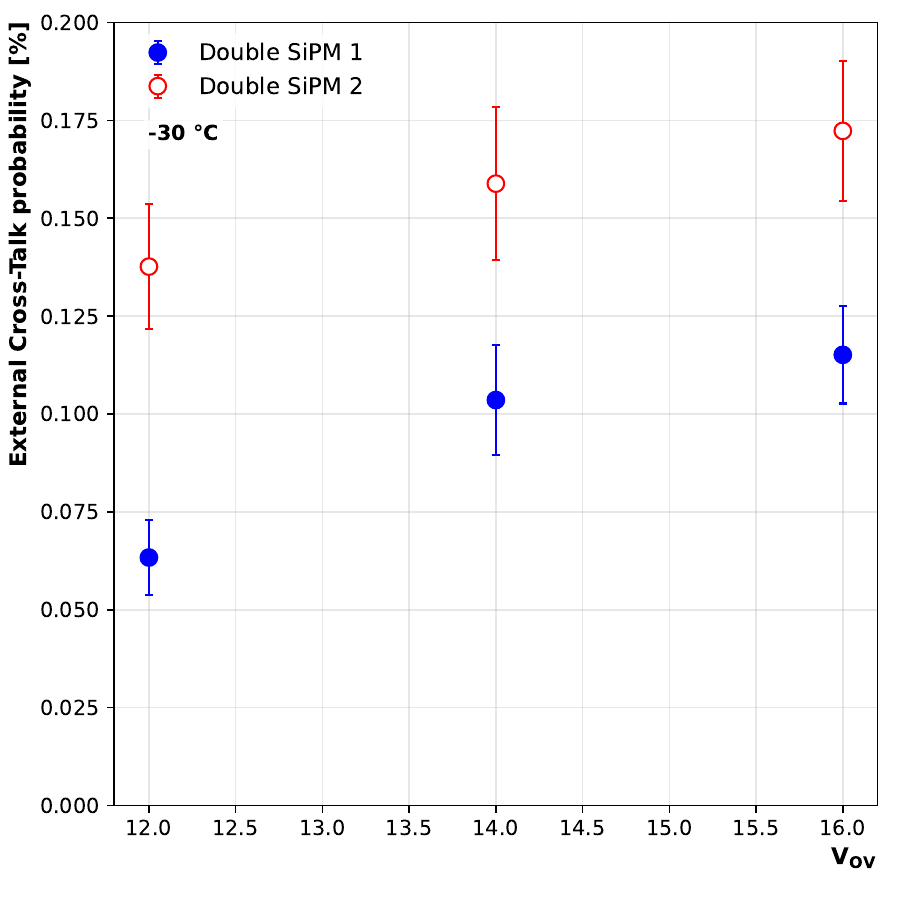}
    \caption{External crosstalk probability as a function of overvoltage at \qty{-30}{\celsius}, for both double SiPMs. The probability increases with overvoltage, reflecting the larger avalanche multiplication at higher SiPM bias.}
    \label{fig:exCT_vs_ov}
\end{figure}

The external crosstalk probability is defined as

\begin{equation}
    p_{exCT} = \frac{n_{exCT}}{N},   
\end{equation}

where $n_{exCT}$ is the number of external crosstalk events, and $N$ the total number of pulses identified on the reference channel in the acquisition window. $n_{exCT}$ is obtained with a fit to the $\Delta t$ distribution using a Gaussian signal and a flat background. The narrow width of the Gaussian fit indicates the excellent timing resolution of this SiPM technology, which is estimated to be better than \qty{350}{\pico\second}. This estimation is limited by the sampling rate of the oscilloscope used. The external crosstalk probability as a function of overvoltage at \qty{-30}{\celsius} is presented in Fig.~\ref{fig:exCT_vs_ov} for both double SiPMs. 

Given the larger avalanche multiplication at higher SiPM bias, the external crosstalk probability is expected to increase with overvoltage. Averaging the two SiPMs, $p_{exCT}$ increases from \qty{0.10 +- 0.04}{\percent} to \qty{0.14 +- 0.03}{\percent}  
over the range \qtyrange{12}{16}{\Vov} at \qty{-30}{\celsius}. However, double SiPM 2 consistently shows higher $p_{exCT}$ across the full overvoltage range. For both double SiPMs, the measured values are approximately two orders of magnitude smaller than the internal direct crosstalk probability, reflecting the geometric suppression of photon propagation between neighbouring SiPM devices.

\section{Conclusions}
\label{sec:conclusions}

A systematic characterisation of commercially available Broadcom AFBR-S4N series NUV-MT SiPMs over a temperature range from \qty{-85}{\celsius} to \qty{23}{\celsius} and operating voltages of \qtyrange{10}{16}{\Vov} has been presented, and complements recent related studies. 

The breakdown voltage, gain, and signal-to-noise ratio exhibit the expected dependences on temperature and overvoltage, in good agreement with manufacturer specifications. Among the noise sources characterised, the dark noise increases strongly with increasing temperature, consistent with thermally activated carrier generation, whilst the afterpulsing probability increases strongly at lower temperatures due to the Shockley--Read--Hall carrier release mechanism. The direct crosstalk probability is found to be approximately independent of temperature but increases with overvoltage, consistent with its avalanche-driven origin. The delayed crosstalk probability increases strongly with decreasing temperature, due to extended carrier diffusion lengths at lower temperatures. These qualitative behaviours are consistent with those reported for other SiPM technologies and, where comparisons are possible, measured parameters are typically in agreement with manufacturer specifications.

External crosstalk between neighbouring channels of the dual-channel model has been characterised using a dedicated timing correlation method in dark conditions. Unlike previous measurements of external crosstalk in NUV-MT devices, which were performed in scintillator-coupled configurations where the result is strongly dependent on detector geometry, the measurements in this study isolate the intrinsic device-level external crosstalk arising from the fixed channel separation within the SiPM package. This provides a geometry-independent lower bound on the external crosstalk probability relevant to any detector configuration employing this SiPM series, and is found to be approximately two orders of magnitude smaller than the internal direct crosstalk probability across all operating conditions investigated.

\section*{Acknowledgements}
The support of the Deutsche Forschungsgemeinschaft (DFG, German Research Foundation) under Germany’s Excellence Strategy – EXC 2121 “Quantum Universe”-390833306 is acknowledged. 

\bibliographystyle{elsarticle-num}
\bibliography{references}   
\end{document}